\documentclass[%
 reprint,
 amsmath,amssymb,
 aps,
]{revtex4-1}

\usepackage{graphicx}
\usepackage{dcolumn}
\usepackage{bm}

\begin{document}

\preprint{This line only printed with preprint option}

\title{Multi-partite quantum nonlocality and Bell-type inequalities in an infinite-order quantum phase transition of the one-dimensional spin-$\frac{1}{2}$ XXZ chain}

\author{Zhao-Yu Sun,$^*$ Shuang Liu, Hai-Lin Huang, Duo Zhang, Yu-Yin Wu, Jian Xu, Bi-Fu Zhan, Hong-Guang Cheng}

\affiliation{School of Electrical and Electronic Engineering, Wuhan Polytechnic
University, Wuhan 430000, China.}

\author{Cheng-Bo Duan}
\affiliation{College of Science, Chang'an University, Xi'an 710064, China}

\author{Bo Wang}
\affiliation{ENN Group Co., Ltd., Langfang, Hebei 065001, China}

\begin{abstract}
In this paper, combined with infinite time-evolving block decimation (iTEBD) algorithm and Bell-type inequalities, we investigate
multi-partite quantum  nonlocality
in an infinite one-dimensional quantum spin-$\frac{1}{2}$ XXZ system.
High \textit{hierarchy} of multipartite nonlocality can be observed in the gapless phase of the model,
meanwhile only the lowest \textit{hierarchy} of multipartite nonlocality is observed in most regions of the gapped anti-ferromagnetic phase.
Thereby, Bell-type inequalities disclose different correlation structures in the two phases of the system.
Furthermore,
at the infinite-order QPT (or Kosterlitz-Thouless QPT) point of the model, the correlation measures always show a local minimum value, regardless of the length of the subchains. It indicates that relatively low \textit{hierarchy} of multi-partite nonlocality would be observed at the infinite-order QPT point in a Bell-type experiment. The result is in contrast to the existing results of the second-order QPT in the one-dimensional XY model, where multi-partite nonlocality with the highest \textit{hierarchy} has been observed.
Thus, multi-partite nonlocality provides us an alternative perspective to distinguish between these two kinds of QPTs.
Reliable clues for the existence of tripartite quantum entanglement have also been found.
\end{abstract}
\maketitle

\section{introduction}

Recently, concepts from quantum information theory
has attracted much attention in the field of condensed matter physics.\cite{information,QE_QPT}
On one hand, various ideas have been used to characterize quantum correlation, for example,
the quantum entanglement defined in the entanglement-separability paradigm\cite{QE_QPT}
and the quantum nonlocality defined by the Bell inequalities\cite{Bell_inequalitiesQPTs_XXZ_model,BFV_Topological_QPT,Bell_QPT_models}.
On the other hand, quantum correlation can be classified according to different partitions of the concerned system.\cite{multi_entanglement}
For instance, a system can just be divided into two parts, i.e., part A and part B. Then the correlation between part A and part B can be called bi-partite correlation
of the system. A system can also be divided into multiple parts, and the situations would become very complex.
Because of its direct physical meaning and the simple form,
bipartite quantum correlation has been studied in many quantum models.
For instance, it has been found that the quantum entanglement entropy (a measure of bipartite entanglement for two spin-blocks) shows a peculiar scaling behavior in the vicinity of quantum phase transition (QPT) points in many systems.\cite{QE_QPT}
In addition, it has been realized that a quantum system would present large amount of bipartite quantum entanglement in the vicinity of the QPT point.\cite{QE_QPT,QE1,QE2,QE3}
These measures of quantum correlation offer us an alternative perspective to investigate QPTs in quantum systems.\cite{BOOK_QPT}

Nevertheless, these concepts also bring new challenges to our understanding of the quantum correlation in condensed matter physics.
For example,
since quantum entanglement is widely present in quantum spin systems, one may expect that quantum nonlocality should show similar behavior.
However, it is rather surprising that in various low-dimensional quantum spin chains, quantum nonlocality is just absent in two-site subchains.\cite{Bell_inequalitiesQPTs_XXZ_model,BFV_Topological_QPT,Bell_QPT_models,nonviolation,violation_PRA,Correlation_nonlocalityQPTS_several_systems,Bell_ladder_MPS}
The existence and distribution of quantum nonlocality in these low-dimensional quantum spin systems have not been fully understood.

Quite recently, the definition of quantum nonlocality has been generalized to multi-partite settings with the help of Bell-type
inequalities.\cite{Multi_Bell1,Multi_Bell2,Multi_Bell3,Multi_Bell4}
In multi-partite settings the situation becomes much more complex than in bipartite setting, and a new idea---the \textit{hierarchy} of multi-partite nonlocality---emerges naturally.
For example, bipartite nonlocality can be regarded as the lowest \textit{hierarchy} of multi-partite nonlocality.
Recently, in an one-dimensional  analytically solvable spin XY chain, multi-partite Bell-type inequalities have been
investigated extensively.\cite{Multi_Bell5}
It's found that high-order Bell inequalities would be violated in the vicinity of the QPT point, meanwhile low-order Bell inequalities would be violated in the non-QPT regions. These results show that the QPT of the system is accompanied by high \textit{hierarchy} of multi-partite nonlocality.
Thereby, the multi-partite Bell inequalities increase our knowledge of the second-order QPT in the XY model.

We would like to mention that the study of multi-partite Bell inequalities in condensed matter physics is far from finished.
Firstly, a technical difficulty may emerge when one tries to study multi-partite Bell inequalities in other low-dimensional quantum models.
In fact, except for few analytically solvable models,
it is nontrivial to obtain multi-site correlations in general quantum spin chains.
Secondly, the behavior of multi-partite nonlocality in other novel QPTs, such as the infinite-order phase transition, is still unknown.\cite{Bell_inequalitiesQPTs_XXZ_model,Renormalization_XXZ}
An infinite-order QPT
shows no singularity in any finite derivative of the ground-state energy, thus is usually very difficult to characterize with traditional methods.
Nevertheless, quantum-information properties may provide unexpected information about the QPT. For example, at the infinite-order QPT point of the XXZ model, measures of quantum entanglement show a singularity after enough iterations of quantum renormalization group.\cite{Renormalization_XXZ}
We hope that multi-partite nonlocality may offer us other valuable information about the type of quantum
correlation involved in infinite-order QPTs.

In this paper, we will use multi-partite Bell-type inequalities to study the quantum nonlocality in an one-dimensional spin XXZ model,
where an infinite-order QPT occurs at $\Delta=1$.\cite{Bell_inequalitiesQPTs_XXZ_model,Renormalization_XXZ}
In contrast with the analytically solvable XY model, in the XXZ model the $n$-site reduced density matrix $\hat{\rho}_n$
is difficult to figure out exactly.
We will use infinite time-evolving block decimation (iTEBD) algorithm to express the ground state of the XXZ model as a matrix product state.\cite{TEBD,MPO}
Then, it is very convenient to identify the reduced density matrix $\hat{\rho}_n$ for any concerned subchain. Consequently, we are ready to investigate the multi-partite nonlocality in the density matrix $\hat{\rho}_n$ with the help of Bell-type inequalities.\cite{Multi_Bell5}
The above procedure can be used to study
multi-partite Bell inequalities in general low-dimensional models.

This paper is organized as follows. We review the concept of multi-partite nonlocality and Bell-type inequalities in Sec. II.
The XXZ model and its solution will be present in Sec. III.
Main results are shown in Sec. IV, and a summary will be given in Sec. V.

\section{multi-partite quantum nonlocality and Bell-type inequalities}

\begin{figure}
\includegraphics{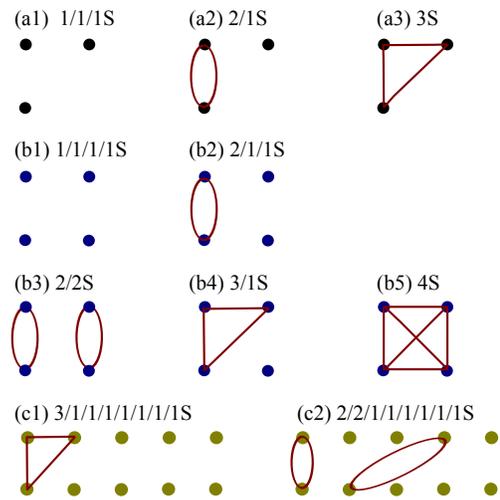}\caption{\label{fig:grouping model}(Color online)
Schematic diagram of various classes of nonlocality for $n$-party systems with $n=3,4$ and $10$.
The solid dots denote the $n$ parties,
and the red lines denote the communication channels between parties.
}
\end{figure}

The definition of multi-partite nonlocality uses the concept of communication from information theory.\cite{Multi_Bell3,Multi_Bell4}
Let's explain the idea by a three-party system($n=3$),
where we will have three classes of nonlocality. (i) Any party  cannot communicate with other parties. It is the so-called
local variables model, and is denoted as 1/1/1S (see Fig. 1(a1)). (ii) Two parties can communicate with each other, and the third one is separated, which is denoted as 2/1S (see Fig. 1(a2)). (iii) All the three parties can communicate with each other, which is denoted as 3S (see Fig. 1(a3)).
There are sufficient criterions to judge which one of the above three classes of nonlocality is present in a three-party system.\cite{Multi_Bell2}
The definition can be easily generalized into large $n$. In Fig. 1(b1)-(b5), we show the five classes of nonlocality in a four-party system($n=4$), denoted as 1/1/1/1S, 2/1/1S, 2/2S, 3/1S, and 4S.
However, in practise it becomes difficult to identify which one of the five classes of nonlocality is present in a four-party system.

An alternative method to quantify multipartite nonlocality
is to use the so-called grouping models, first proposed by Bancal et al.\cite{Multi_Bell3,Multi_Bell4}
Suppose an $n$-party system can be divided
into $k$ groups($1 \le k\le n$), such that every party can just communicate
with others in the same group. It will be called an $(n,k)$-grouping model.
For instance,
$(3,k)$-grouping models with $k=3,2,1$ are equivalent to $\{$1/1/1S$\}$, $\{$2/1S$\}$, $\{$3S$\}$, respectively.
$(4,k)$-grouping models with $k=4,3,2,1$ are corresponding to $\{$1/1/1/1S$\}$, $\{$2/1/1S$\}$, $\{$2/2S,3/1S$\}$, $\{$4S$\}$, respectively.
One can see that $(4,2)$-grouping models contains two sub-classes, i.e., 2/2S and 3/1S.
Similarly, $(10,8)$-grouping models also contains two sub-classes, i.e., 3/1/1/1/1/1/1/1S and 2/2/1/1/1/1/1/1S (please see Fig. 1(c1), (c2)).



As we will show, using grouping models to quantify multipartite nonlocality is convenient in practice.
To make the statement more concise, in this paper, if $(n,k)$-grouping models are needed to reproduce the quantum correlation in the system,
we will just say that the system contains ($n$,$k$)-type nonlocality.
We would like to mention that the two-site quantum nonlocality, which have been widely studied in quantum spin systems with the help of the
Bell-Clauser-Horne-Shimony-Holt (Bell-CHSH) inequality,\cite{Bell_Bell_Inequalityies,CHSH,Horodecki_BFV_twoQubitState}
can be regarded as $(2,1)$-type nonlocality in the framework of \textit{grouping models} theory.
Now it is clear that two-site quantum nonlocality would not be present in most translational invariant systems, such as in the XXZ model.
A further description and characterization of quantum nonlocality of the XXZ model is one of the goals of this paper.

To proceed,
let's consider an $n$-party system described by a density matrix $\hat{\rho}_n$.
We will use
$\vec{a}_j$ and $\vec{a}'_{j}$ with $j=1,...,n$ to denote unit vectors
in $R^{3}$ space,
and $\vec{{\sigma}}=(\hat{\sigma}_{x},\hat{\sigma}_{y},\hat{\sigma}_{z})$
is the spin vector, whose elements are the three Pauli matrices.
We first define $\hat{M}_{1}=\vec{a}_{1}\cdot\vec{\sigma}$ and $\hat{M}'_{1}=\vec{a}'_{1}\cdot\vec{\sigma}$, then the Mermin-Klyshko (MK) operators are defined
recursively as \cite{ineq30,ineq31,ineq32}
\[
\begin{array}{c}
\hat{M}_{n}=\frac{1}{2}\hat{M}_{n-1}\otimes(\vec{a}_{n}+\vec{a}'_{n})\cdot\vec{\sigma}_n+\frac{1}{2}\hat{M}'_{n-1}\otimes(\vec{a}_{n}-\vec{a}'_{n})\cdot\vec{\sigma}_n,\\
\hat{M}'_{n}=\frac{1}{2}\hat{M}'_{n-1}\otimes(\vec{a}'_{n}+\vec{a}{}_{n})\cdot\vec{\sigma}_n+\frac{1}{2}\hat{M}{}_{n-1}\otimes(\vec{a}'_{n}-\vec{a}{}_{n})\cdot\vec{\sigma}_n.
\end{array}
\]


In order to detect ($n$,$n$-$m$)-type nonlocality with $m=1,3,5,7...$,
we need to consider the $m$-order Mermin inequality
\begin{equation}
M_{n}=\textrm{Tr}(\hat{\rho}_{n}\hat{M}_{n})\le 2^{(m-1)/2}\label{eq:Inequality_entanglement}.
\end{equation}
If inequality (\ref{eq:Inequality_entanglement}) is violated,
one can prove that the system contains ($n$,$n$-$m$)-type nonlocality.
In practice, one should figure out the maximum value of $M_{n}$ by optimizing $\textrm{Tr}(\hat{\rho}_{n}\hat{M}_{n})$ with respect to all the unit vectors  \{$\vec{a_{1}}$,$\vec{a}'_{1}$,...$\vec{a_{n}}$,$\vec{a}'_{n}$\}.

On the other hand, when $m=2,4,6,8...$, let's denote $\hat{M}_{n+}=\frac{1}{\sqrt{2}}(\hat{M}_{n}+\hat{M}'_{n})$,
and define the following $m$-order Mermin-Svetlichny inequality \cite{Multi_Bell4}
\begin{equation}
M_{n+}=\textrm{Tr}(\hat{\rho}_{n}\hat{M}_{n+})\le 2^{(m-1)/2}.\label{eq:inequality_nonlocality}
\end{equation}
If inequality (\ref{eq:inequality_nonlocality}) is violated,
we conclude that $\hat{\rho}_{n}$ contains ($n$,$n$-$m$)-type nonlocality.

Inequalities (\ref{eq:Inequality_entanglement}) and (\ref{eq:inequality_nonlocality})
are exclusively used to identify whether or not a multi-party state contains nonlocal correlation.
In some situations, the quantities $M_n$ and $M_{n+}$ cannot
reach a value sufficient to violate the inequalities.
As we will show, however, the functional behavior in $M_n$ and $M_{n+}$ is also informative.


\section{model and solution}

We consider an one-dimensional infinite spin-$\frac{1}{2}$ XXZ model described by the
following Hamiltonian\cite{Bell_inequalitiesQPTs_XXZ_model,BOOK_QPT,Correlation_nonlocalityQPTS_several_systems,Renormalization_XXZ}

\[
\hat{H}=\sum_{i}\sigma_{i}^{x}\sigma_{i+1}^{x}+\sigma_{i}^{y}\sigma_{i+1}^{y}+\Delta\sigma_{i}^{z}\sigma_{i+1}^{z},
\]
where $\sigma_{i}^{\alpha}$ with $\alpha=x,y,z$ denote the Pauli
matrices on site $i$, and $\Delta$ is the anisotropic parameter.
We consider a periodic boundary condition with $\sigma_{i+N}^{\alpha}=\sigma_{i}^{\alpha}$.
The model has three phases. In the limit $\Delta\rightarrow+\infty$,
the interactions in the $x$ and $y$ directions will be ignored,
thus the model should be in an anti-ferromagnetic phase. On the other
hand, in the limit $\Delta\rightarrow-\infty$, the model should be
in a ferromagnetic phase. In the intermediate region the system is
in a gapless phase, and there is a first-order QPT at $\Delta=-1$
and an infinite-order QPT at $\Delta=+1$.


Our goal is to investigate multi-partite quantum correlation in the infinite-order QPT of the model.
We would like to mention that the first-order QPT at $\Delta=-1$ has been studied by considering tripartite entanglement
in a very recent paper.\cite{XXZ_first_order_QPT}

Though the ground state
energy of the XXZ model can be obtained easily with the Bethe ansatz
method,\cite{Yang} the multi-spin reduced density matrix is difficult to calculate
exactly. Thereby, in this paper, we propose the following numerical procedure to investigate
the multi-partite nonlocality in the model. The framework
can be used to study general one-dimensional quantum spin systems.

(1) Firstly, we will use the iTEBD algorithm to express approximately the ground state $|\psi_g\rangle$
of the infinite XXZ model by a matrix product state (MPS)\cite{TEBD}

\[
|\psi_{mps}\rangle=\sum_{i_{1}i_{2}i_{3}...i_{N}}\textrm{Tr}(A_{i_{1}}A_{i_{2}}A_{i_{3}}\cdots A_{i_{N}})|i_{1}i_{2}i_{3}...i_{N}\rangle,
\]
where $\{A_{i}\}$ are $D\times D$ matrices, and $i_j=0,1$ denote the spin-down state and the spin-up state of the $j$-th spin. $D$ is
the dimension of the MPS, which controls the accuracy of the algorithm.
The basic idea of the iTEBD algorithm is that, for an arbitrary initial
state ($|\psi_{ini}\rangle$) which has non-zero overlap with the
ground state $|\psi_{g}\rangle$, the system will evolve to the ground
state if we carry out imaginary-time evolution repeatedly,
i.e., $(e^{-\tau\hat{H}})^{M}|\psi_{ini}\rangle\longrightarrow|\psi_{g}\rangle$,
where $\tau$ is the quantity defined in the Trotter-Suzuki
decomposition, and $M$ denotes the steps of the imaginary-time
evolution.
With a small enough $\tau$, the operator $e^{-\tau\hat{H}}$ is decomposed into a sequence of two-site operators $e^{-\tau\hat{H}_{i,i+1}}$.
To carry out a single step of imaginary-time evolution $e^{-\tau\hat{H}}|\psi_{in}\rangle\longrightarrow|\psi_{out}\rangle$ efficiently,
we need to express the input wave-function as an MPS and the two-site operator $e^{-\tau\hat{H}_{i,i+1}}$ as an matrix product operator (MPO).
More details about the iTEBD algorithm can be found in Refs. \cite{TEBD,MPO}.
In our calculations, the dimension of the MPS is set as $D=16$.
In order to check the accuracy of the method,
we have compared the approximate ground-state energy $E_{mps}=\langle\psi_{mps}|\hat{H}|\psi_{mps}\rangle$
with the exact ground-state energy $E_{exact}$.
The relative error $\frac{E_{mps}-E_{exact}}{E_{exact}}$
turns out to be smaller than $1.6\times10^{-4}$ for any $\Delta$.

We would like to mention that the iTEBD algorithm is an alternative
procedure of the famous density matrix renormalization group (DMRG)
method.\cite{DMRG}
The original DMRG method is very powerful in calculating the
reduced density matrix $\tilde{\rho}_{n}$ of very large blocks. However,
the obtained $\tilde{\rho}_{n}$ by DMRG method is expressed in a
truncated Hilbert space. The Bell-type inequalities are just defined
in the standard Hilbert space. Thus the iTEBD algorithm rather than the DMRG method is adopted in this paper.

(2) Secondly, we will calculate the reduced density matrix $\hat{\rho}_{n}$
of the continuous $n$-site subchain. One will find that it is very
convenient to identify $\hat{\rho}_{n}$ from an MPS, that is,
the elements of $\hat{\rho}_{n}$ can be calculated directly from $\{A_{i}\}$ as
\begin{equation}
\rho_{i_{1},...,i_{n},j_{1},...,j_{n}}=\langle\lambda\vert{A}_{i_{1}}^{*}...{A}_{i_{n}}^{*}\otimes{A}_{j_{1}}...{A}_{j_{n}}\vert\lambda\rangle.\label{eq:rho_general}
\end{equation}
where ${A}_{i}^{*}$ is the complex conjugate of ${A}_{i}$, and
$\lambda$ is the largest eigenvalue of the transfer matrix $T=\sum_i {A}_{i}^{*} \otimes{A}_{i}$, with
 $\langle\lambda\vert$ and $\vert\lambda\rangle$ the corresponding left and right eigenvectors.

(3) Thirdly, we numerically optimize the value
of $\textrm{Tr}(\hat{\rho}_{n}\hat{M}_{n})$ (and $\textrm{Tr}(\hat{\rho}_{n}\hat{M}_{n+})$)
with respect to all the unit vectors $\vec{a}_{j}=(\sin\theta_j\cos\phi_j,\sin\theta_j\sin\phi_j,\cos\theta_j)$,
and resort to the inequalities (\ref{eq:Inequality_entanglement}) and (\ref{eq:inequality_nonlocality}) to analyze the multi-partite
quantum nonlocality in $\hat{\rho}_{n}$.
The numerical optimization is very time-consuming.
Some numerical details
for improving the efficiency of multi-variable numerical optimization can be
found in Refs. \cite{Multi_Bell5,Camp}.
In this paper, when the length of the subchain is short ($n\le4$), we are able to optimize the value of $\textrm{Tr}(\hat{\rho}_{n}\hat{M}_{n})$
with respect to $\vec{a}_{j},\vec{a}'_{j}\in R^3$ for $j=1,...,n$.
When $n$ is large ($n>4$), we just consider two special situations;
that is, all the unit vectors
$\vec{a}_{j}$($\vec{a}'_{j}$) are constrained in the $x$-$y$ plane, or in the $x$-$z$ plane.\cite{ZZZZ}
We will use $M_n$, $M_n^{xy}$ and  $M_n^{xz}$ to denote the
optimized results according to the entire $R^3$ space, the $x$-$y$ plane, and the $x$-$z$ plane, respectively.
For $n\le4$, we numerically find that $M_n=\max\{M_n^{xy},M_n^{xz}\}$ and $M_{n+}=\max\{M_{n+}^{xy},M_{n+}^{xz}\}$
for any $\Delta$.

\section{multi-partite nonlocality in the XXZ model}
In subsection \ref{sebsection:1}, we will study the lowest \textit{hierarchy} of quantum nonlocality, that is,
($n$,$n$-1)-type nonlocality, by the Mermin
inequality (\ref{eq:Inequality_entanglement}). Then in subsection \ref{sebsection:2},
($n$,$n$-2)-type quantum nonlocality will be studied by the Mermin-Svetlichny  inequality (\ref{eq:inequality_nonlocality}).

\begin{figure}
\includegraphics{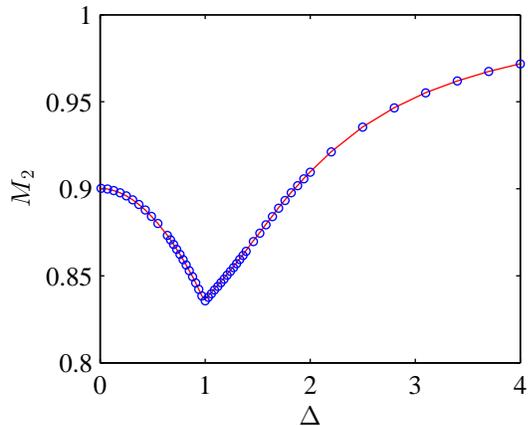}\caption{\label{fig:M2}(Color online) Blue circles denote the violation measure $M_2$ of the Mermin inequality. The red solid line is according to
Horodecki¡¯s formula, divided by 2.}
\end{figure}

\begin{figure}
\includegraphics{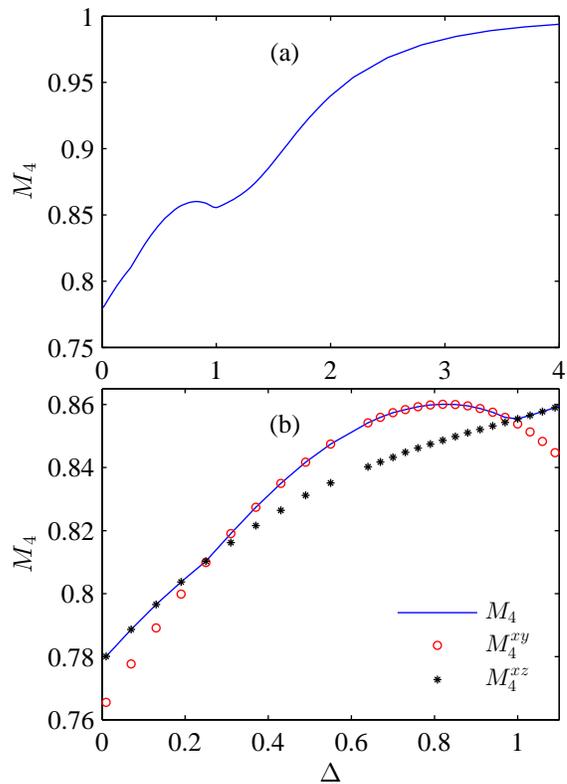}\caption{\label{fig:M4}(Color online)
(a) The violation measure $M_4$ for a 4-site subchain in the XXZ model.
(b) The violation measures $M_4$ (blue line), $M_4^{xy}$(red circles) and $M_4^{xz}$(black stars) in the vicinity of $\Delta=1$.
}
\end{figure}

\begin{figure}
\includegraphics{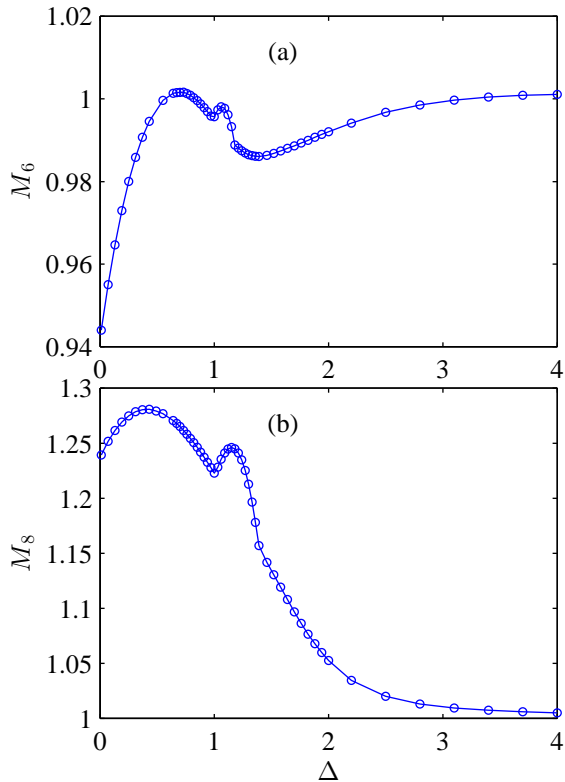}\caption{\label{fig:M6}(Color online)
The violation measure $M_n$ for (a) a 6-site subchain and (b)  an 8-site subchain in the XXZ model.
}
\end{figure}

\subsection{($n$,$n$-1)-type nonlocality}\label{sebsection:1}
We try to detect multi-partite nonlocality in $n$-site subchains by considering the first non-trivial inequality, i.e.,
the $1$-order Mermin inequality,
\begin{equation}
M_{n}=\textrm{Tr}(\hat{\rho}_{n}\hat{M}_{n})\le 1.\label{eq:1}
\end{equation}
According to Sec. II, the violation of the above inequality indicates that the subchain described by $\hat{\rho}_n$
contains ($n$,$n$-$1$)-type nonlocality.

Firstly, we consider a two-spin subchain, i.e., $n=2$. In addition to numerical optimization, the
value of $M_2$ can also be obtained exactly with Horodecki's formula.\cite{Horodecki_BFV_twoQubitState}
Our results are shown in Fig. \ref{fig:M2}.
One can see that the results from our numerical optimization (see blue circles) are in good agreement with the results from Horodecki's formula (see the red line),
thus the validity of our procedure of numerical optimization is confirmed.
Furthermore, the inequality (\ref{eq:1}) is not violated in the whole parameter space.
It means that ($2$,$1$)-type quantum nonlocality has not been observed.
It is interesting that $M_2$ shows a sharp bottom at $\Delta=1$, i.e., the infinite-order QPT point of the system.
The above results are consistent with previous studies using the CHSH inequality to study the two-site quantum nonlocality in the XXZ
model.\cite{Bell_inequalitiesQPTs_XXZ_model,Correlation_nonlocalityQPTS_several_systems}

We make a further step to consider a four-site subchain and the results are shown in Fig. \ref{fig:M4}(a). The inequality (\ref{eq:1}) is not violated for any $\Delta$, thus (4,3)-type nonlocality is not observed in the subchain.
However, we find that $M_4$ again shows a valley at the QPT point $\Delta=1$.
It would be interesting that whether or not $M_n$ always shows a valley at the QPT point for larger $n$.

For $n\le4$, we numerically found that $M_n=\max\{M_n^{xy},M_n^{xz}\}$ (see Fig. \ref{fig:M4}(b) for $n=4$). Thus,
in order to deal with  $M_6$ and $M_8$, we will constrain all the unit vectors in the $x$-$y$ plane and the $x$-$z$ plane.

Fig. \ref{fig:M6}(a) shows the results for $n=6$. As the increase of $\Delta$, $M_6$ shows a bimodal shape,
and the valley between the two peaks locates at the QPT point $\Delta=1$.
Moreover, we find that in the vicinity of the peak point $\Delta\approx0.7$, the value of $M_6$ is slightly larger than 1
, thus
(6,5)-type quantum nonlocality is observed in the system.
It is unexpected that the quantum nonlocality is observed at $\Delta\approx0.7$, rather than at the QPT point,
since one usually expects that a system would have stronger quantum correlation in the QPT than in non-QPT regions.


The results for $n=8$ are shown in Fig. \ref{fig:M6}(b). The bimodal shape is observed again, with the valley locating at the  QPT point $\Delta=1$.
In addition, in the whole parameter space $\Delta>0$, the value of $M_8$ is always larger than 1, thus the inequality (\ref{eq:1}) is violated.
It shows that (8,7)-type quantum nonlocality is always present in the $8$-site subchain.


Previous studies show that (2,1)-type nonlocality (two-site nonlocality)
is not present in most one-dimensional spin chains,\cite{nonviolation,violation_PRA}
including the XXZ model\cite{Bell_inequalitiesQPTs_XXZ_model}.
Our results show that ($n$,$n$-1)-type nonlocality indeed exists in the XXZ model. However, it is not shared by any two sites, instead,
it is distributed in long subchains in the form of multi-site quantum nonlocality.

\begin{figure}
\includegraphics{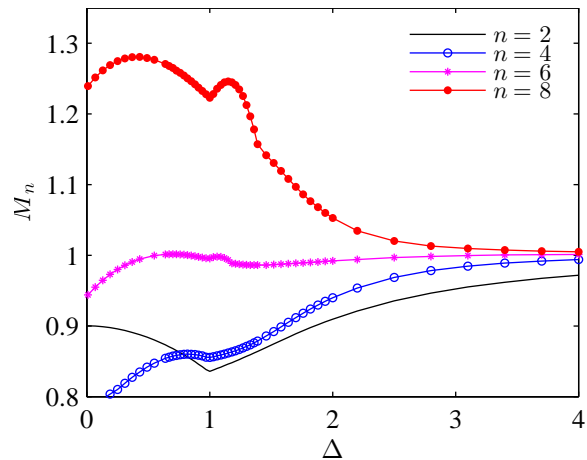}\caption{\label{fig:Mn2}(Color online)
The violation measure $M_{n}$ for $n$-site subchains in the XXZ model.
}
\end{figure}

One may further consider ($n$,$n$-$m$)-type nonlocality with $m=3,5,7   ...$.
For this purpose, we need to consider the $m$-order Mermin inequalities
$M_{n}\le 2, 4, 8...$, respectively.
We combine our results of $M_{n}$ for $n=2,4,6,8$ in Fig. \ref{fig:Mn2}.
The above inequalities are never violated in the whole parameter space,
however, we can make some reliable evaluation.
In the far-from critical regions of the anti-ferromagnetic phase, as the increase of $n$, $M_n$ converges quickly.
Consequently, the above inequalities would never be violated and ($n$,$n$-$m$)-type nonlocality with $m=3,5,7...$ should never be observed.
On the other hand,
in the gapless phase $\Delta\in[0,1]$ and the near-critical regions of the anti-ferromagnetic phase($\Delta\gtrsim1$),
$M_{n}$ increases steadily, thus the above inequalities may be violated when $n$ is large enough.

\subsection{($n$,$n$-2)-type nonlocality}\label{sebsection:2}
\begin{figure}
\includegraphics{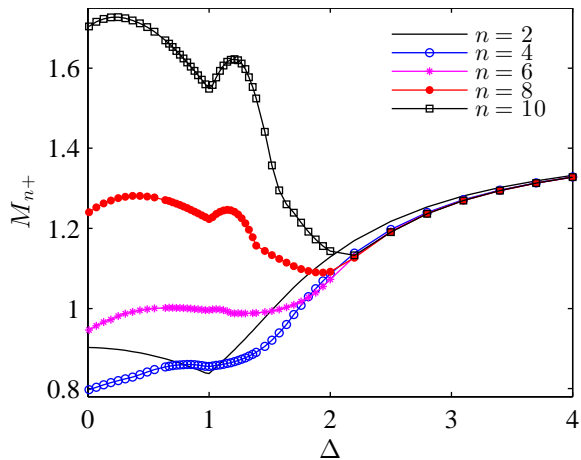}\caption{\label{fig:Mn+2}(Color online)
The violation measure $M_{n+}$ for $n$-site subchains in the XXZ model.
}
\end{figure}

Let's consider a high \textit{hierarchy} of quantum nonlocality with an even $m$, i.e., ($n$,$n$-$2$)-type nonlocality.
We need to use the $2$-order Mermin-Svetlichny inequality (\ref{eq:inequality_nonlocality}), i.e.,
\begin{equation}
M_{n+}=\textrm{Tr}(\hat{\rho}_{n}\hat{M}_{n+})\le \sqrt{2}.\label{eq:3}
\end{equation}
For $n\le4$ we numerically find that the value of $M_{n+}$ is exactly equal to $\max\{ M_{n+}^{xy},M_{n+}^{xz}\}$. Thus we will constrain all the unit vectors
in the $x$-$y$ plane and the $x$-$z$ plane, just as in calculating $M_n$. We combine the results of $M_{n+}$ for several $n$ in Fig. \ref{fig:Mn+2}.

\begin{figure}
\includegraphics{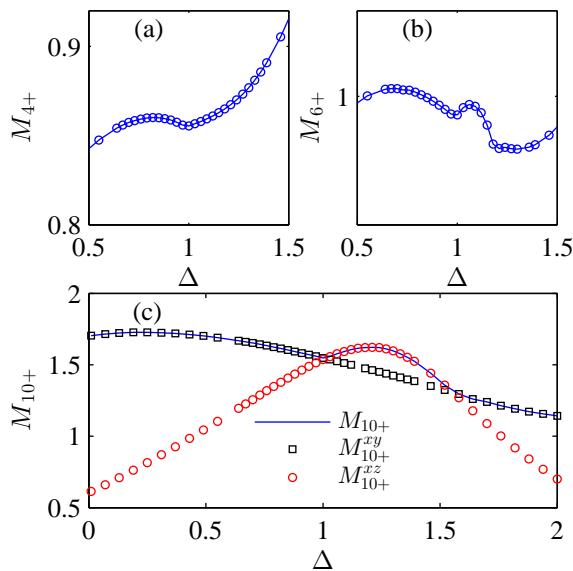}\caption{\label{fig:Mn+1}(Color online)
The violation measure $M_{n+}$ for $n$-site subchains in the XXZ model.
A size-independent valley is observed at the finite-order QPT point $\Delta=1$.
}
\end{figure}

Firstly, let's consider the far-from critical region in the anti-ferromagnetic phase,
i.e., $\Delta\gg1$.
It is clear that as the increase of $n$, $M_{n+}$ converges very fast. In fact, we can not distinguish between the curves of $M_{6+}$, $M_{8+}$ and $M_{10+}$ from each other.
As a result, the inequality $M_{n+}\le\sqrt{2}$  should never be violated,
regardless of the length of the subchain.
Next, we consider the gapless phase $\Delta\in[0,1]$ and the near-critical region in the anti-ferromagnetic phase ($\Delta\gtrsim1$).
For $n\le8$, the inequality $M_{n+}\le\sqrt{2}$ is never violated. However,
as the increase of $n$,
$M_{n+}$ shows a clear non-convergence behavior, and inequality is finally be violated for $n=10$.
Thereby, ($n$,$n$-$2$)-type nonlocality is observed.
As we have discussed in Sec. II, (10,8)-type nonlocality contains two sub-classes,
as illustrated in Fig. \ref{fig:grouping model}(c1) and (c2).
It is still unknown how to identify which sub-class of nonlocality is involved in the system.

One may further detect ($n$,$n$-$m$)-type nonlocality with $m=4,6,8$...by considering the $m$-order Mermin-Svetlichny inequalities
$M_{n+}\le 2\sqrt{2}, 4\sqrt{2}, 8\sqrt{2}...$.
In the gapless phase $\Delta\in[0,1]$ and the near-critical regions of the anti-ferromagnetic phase($\Delta\gtrsim1$),
since $M_{n+}$ increases steadily, the above inequalities may be violated when $n$ is large enough.
In the far-from critical regions of the anti-ferromagnetic phase($\Delta\gg1$), since $M_n$ converges quickly as the increase of $n$,
the above inequalities should never be violated and ($n$,$n$-$m$)-type nonlocality with $m=4,6,8...$ should never be observed.
Combined with the results in Sec. IV A, we conclude that when $\Delta\gg1$,
quantum nonlocality should just be distributed in the form of ($n$,$n$-$1$)-type nonlocality, rather than any other high \textit{hierarchy} of multipartite nonlocality.
One can see that in the gapless phase and the anti-ferromagnetic phase, the system shows different quantum correlation structures,
that is,
high \textit{hierarchy} of multipartite nonlocality would be observed in the gapless phase, meanwhile only the ($n$,$n$-$1$)-type nonlocality can be observed in most regions of the anti-ferromagnetic phase.

We pay our attention to the infinite-order QPT point $\Delta=1$.
It is clear that both $M_n$ and $M_{n+}$ always show a valley at the QPT point, regardless of the length of the subchain.
Details for $M_{4+}$ and $M_{6+}$ can be found in Fig. \ref{fig:Mn+1} (a) and (b).
In fact, the first-order derivative of $M_{n}$ (and $M_{n+}$) is discontinuous at $\Delta=1$.
This singularity results from the sudden change of the panel in which the optimal $M_{n}$($M_{n+}$) is obtained,
in other words, the curves of $M_{n}^{xy}$($M_{n+}^{xy}$) and $M_{n}^{xz}$($M_{n+}^{xz}$) cross each other at $\Delta=1$.
For example, see Fig. \ref{fig:M4}(b) for $M_n$  and Fig. \ref{fig:Mn+1}(c) for $M_{n+}$.
Similar behavior has been reported in a recent paper by S. Campbell\cite{Camp},
in which the optimal angle for calculating quantum global discord of the transverse-field Ising model also changes suddenly in the vicinity of the QPT point.
The sharp change of the optimal panels(angles) may be related to the dramatic change of the wave-function at the QPT point.
In addition, we find that the curves of $M_{n}^{xy}$($M_{n+}^{xy}$) and $M_{n}^{xz}$($M_{n+}^{xz}$) also cross each other at non-QPT regions. As a result, an additional singularity is observed at some $\Delta\ne1$.
Please see the point $\Delta\approx0.3$ in Fig. \ref{fig:M4}(b) and  $\Delta\approx1.5$ in Fig. \ref{fig:Mn+1}(c).
According to Figs. \ref{fig:Mn2} and \ref{fig:Mn+2}, this additional singularity has a clear dependence upon the length of the subchain.
Thus its behavior is fundamentally different from the size-independent singularity at the QPT point.
We would like to mention that this kind of discontinuity can also be observed in any other physical quantity when max or min function is involved in its definition,
for instance, the entanglement concurrence (defined by a max function) and the quantum discord (defined by a min function).
The singularity of concurrence appearing in non-QPT regions has been discussed in Ref. \cite{CC}.
A detailed analysis about the singularity of discord can be found in Ref. \cite{DD}.

Furthermore, the minimum value of $M_{n}$ and $M_{n+}$ at $\Delta=1$ discloses some features of quantum correlation in the QPT.
According to Eqs. (\ref{eq:Inequality_entanglement}) (and (\ref{eq:inequality_nonlocality})),
the value of $M_n$ ($M_{n+}$) determines the order of Mermin (Mermin-Svetlichny) inequality which can be violated.
When $n\rightarrow+\infty$, on the two sides of the QPT point, the value of $M_n$ ($M_{n+}$) would be very large thus very high-order Mermin (Mermin-Svetlichny) inequalities may be violated,
meanwhile at the QPT point, the value of $M_n$ ($M_{n+}$) would be relatively small thus lower order of Mermin (Mermin-Svetlichny) inequalities would be violated.
Then it is expected that, when the system evolves towards to the QPT point,
the \textit{hierarchy} of multi-partite nonlocality will decrease gradually,
and the system would show relatively low \textit{hierarchy} of multi-partite nonlocality at the QPT point.
An opposite behavior has been reported in the second-order QPT of the XY model, where the highest \textit{hierarchy} of multi-partite nonlocality is observed at the QPT point.\cite{Multi_Bell5}
Thereby, multi-partite nonlocality provides us an interesting perspective to distinguish between
these two QPTs.

\section{discussions and summary}

In this paper, combined with iTEBD algorithm and Bell-type inequalities,
we have investigated the multi-partite nonlocality
of continuous $n$-site subchains in an infinite one-dimensional quantum spin XXZ model.
Different quantum correlation structures have been found in the two condensed matter phases of the model.
In the gapless phase, high \textit{hierarchy} of multipartite nonlocality (at least ($n$,$n$-2)-type nonlocality) can been observed,
meanwhile in most regions of the anti-ferromagnetic phase,
merely ($n$,$n$-1)-type nonlocality is observed.
Thus, Bell-type inequalities have increased our understanding of condensed matter phases of the system.

Furthermore, at the infinite-order QPT point $\Delta$=$1$, a size-independent local minima has been
observed in both $M_n$ and $M_{n+}$.
As infinite-order QPTs are usually more difficult to identify than first- and second- order QPTs,
this kind of size-independent minima may be valuable in detecting infinite-order QPTs.
It would be interesting to investigate if this result is only specific to the XXZ model, or is general in other infinite-order QPTs.

Because of the local minima of $M_n$ and $M_{n+}$ at $\Delta$=$1$,
relatively low \textit{hierarchy} of multi-partite quantum correlation will be observed at the QPT point.
For example, when $n=6$, ($n$,$n$-1)-type nonlocality is observed in the vicinity of $\Delta\approx0.7$, rather than at the QPT point.
In the limit $n\rightarrow+\infty$,
it is expected that, when the system evolves away from (towards to) the phase transition point,
the \textit{hierarchy} of multi-partite nonlocality will increase (decrease) gradually.
An opposite result has been reported in the second-order QPT of the XY model.\cite{Multi_Bell5}
Thus, multi-partite nonlocality provides us an interesting perspective to distinguish between
these two QPTs.

The Bell-type inequalities used in this paper is a generalization of the CHSH inequality.
The CHSH inequality has been used to study two-site quantum nonlocality in various models,
such as the XXZ model.\cite{Bell_inequalitiesQPTs_XXZ_model}
The mathematic singularity of the violation measure of the CHSH inequality has already been
used to characterize QPTs in these models. However, since the CHSH inequality would
not be violated in most translation-invariant systems,\cite{nonviolation,violation_PRA}
a high value or a low value of the violation measure does not have specific physical meaning.
However, in Bell-type inequalities, the value of the violation measures ($M_n$ and $M_{n+}$) is directly related to
the order of the inequalities which would be violated.
Thus, the advantage of Bell-type inequalities is that, they characterize QPTs not only by the singularity
but also by the \textit{hierarchy} of multi-partite nonlocality. As a result,
in the two condensed matter phases of the model, different quantum correlation structures have been observed,
and the result is far beyond the result obtained by the CHSH inequality.

We would like to mention that, Bell-type inequalities can also be used as
sufficient criterions to study multipartite quantum entanglement.
A useful introduction of multipartite quantum entanglement can be found in Refs.\cite{multi_entanglement,tau}.
For an $n$-party system with at most $m$-partite entanglement,
one can prove that
$M_n=\textrm{Tr}(\hat{\rho}_{n}\hat{M}_{n})\le 2^{(m-1)/2}$.\cite{Multi_Bell2}
If the value of $M_n$ turns out to be larger than $2^{(m-1)/2}$, we can conclude that there is at least ($m$+1)-partite entanglement
in $\hat{\rho}_{n}$. For instance, we can identify the existence of tripartite entanglement
if $M_{n}\le\sqrt{2}$ is violated.
In Fig. \ref{fig:Mn2}, the value of $M_n$ is always smaller than $\sqrt{2}$ for $n\le8$.
However, because of the non-convergence behavior of $M_n$ when  $\Delta\le1$ and $\Delta\gtrsim1$,
we believe that tripartite entanglement should be observed
when $n$ is large enough.

The procedure proposed in this paper can be used to study  multi-partite nonlocality in other
low-dimensional infinite-size systems, and work is far from finished.
As we have shown, an ($n$,$m$)-type nonlocality may contains various subclasses of nonlocality.
It is still unknown how to identify which sub-class of nonlocality is involved in a quantum system for $n>3$.
Thus, further investigations are needed to disclose more in-depth features of multipartite nonlocality
in one-dimensional quantum spin chains.
Moreover, there are also some other powerful methods to study multi-partite correlation,\cite{tau,Camp}
which may be valuable to characterize the QPT in the XXZ model.

\section*{Acknowledgments}
The research was supported by the National Natural Science Foundation
of China (Nos. 11204223, 11404250 and 61404095). This work was also supported by the Talent
Scientific Research Foundation of Wuhan Polytechnic University (Nos.
2011RZ15, 2012RZ09, 2014RZ18).



\begin{thebibliography}{References}
\bibitem{ZZZZ} Because of the symmetry of the XXZ model, the $y$-$z$ plane will be equivalent to the $x$-$z$ plane.
\end{thebibliography}
\end{document}